\documentclass[preprint,12pt]{elsarticle}

\usepackage{amssymb}
\usepackage{changes}
\usepackage{changes}
\makeatletter
\AddToHook{cmd/added/before}{\def\Changes@AuthorColor{orange}}
\AddToHook{cmd/deleted/before}{\def\Changes@AuthorColor{red}}
\AddToHook{cmd/replaced/before}{\def\Changes@AuthorColor{orange}}
\makeatother
  \usepackage{lineno}
\usepackage{amsmath}
\usepackage{graphicx}
\usepackage{siunitx}
\usepackage{multicol}
\usepackage{todonotes}
\usepackage{placeins}
\usepackage{hyperref}
\hypersetup{
    colorlinks=true,
    linkcolor=blue,
    citecolor=blue,
    filecolor=magenta,      
    urlcolor=cyan,
    }

\journal{Nuclear Instruments and Methods A}
\begin{document}
\begin{frontmatter}


\title{A Novel, Steerable, Low-Energy Proton Source for Detector Characterization}



\author[UofM]{Nicholas Macsai} 
\author[UofM]{August Mendelsohn} 
\author[UofM]{David Harrison}
\author[UofW,UofM]{Russell Mammei} 
\ead{r.mammei@uwinnipeg.ca}
\author[UofM]{Michael Gericke}
\author[ORNL]{Leah Broussard }
\author[LANL]{Erick Smith}
\author[LANL]{Grant Riley}
\author[UofA]{Glenn Randall}
\author[LANL]{Mark Makela}


\affiliation[UofM]{organization={Department of Physics and Astronomy},
             addressline={University of Manitoba}, 
             city={Winnipeg},
             postcode={R3T 2N2}, 
             state={Manitoba},
             country={Canada}
             }

\affiliation[UofW]{organization={Department of Physics},
            addressline={University of Winnipeg}, 
            city={Winnipeg},
            postcode={R3B 2E9}, 
            state={Manitoba},
            country={Canada}
            }

\affiliation[ORNL]{organization={Physics Division},
            addressline={Oak Ridge National Laboratory}, 
            city={Oak Ridge},
            postcode={37831}, 
            state={Tennessee},
            country={USA}
            }
\affiliation[UofA]{organization={Department of Physics},          
            addressline={Arizona State University}, 
            city={Tempe},
            postcode={85287}, 
            state={Arizona},
            country={USA}
            }

\affiliation[LANL]{organization={Los Alamos National Laboratory},          
            city={Los Alamos},
            postcode={87545}, 
            state={New Mexico},
            country={USA}
            }




 


\begin{abstract}
We report on the conversion of the Manitoba II mass spectrometer into a versatile low-energy proton beam facility. 
This infrastructure is adaptable to any detector-under-test (DUT), and has proven itself effective with the characterization of silicon detectors used in subatomic beyond-the-Standard-Model (BSM) searches, namely the Nab experiment. 
A pencil beam of mono-energetic protons can be produced in a range from \SI{25}{\kilo e\volt} to \SI{35}{\kilo e\volt}, achieving a beam current of $\sim$\SI{1e-18}{\ampere}. 
Electrostatic steering plates were constructed to direct the Gaussian-profile proton beam over a \SI{117}{\milli\meter} diameter area-of-interest with full-width at half-maxima (FWHM) ranging from \SI{0.6}{\milli\meter} to \SI{1.26}{\milli\meter}.   
This work discusses the modifications and subsequent tests to confirm the beam specifications met the demands of the aforementioned detectors.


\end{abstract}






\end{frontmatter}

\section{Introduction}
\label{sec:protonsource}

Low energy, high precision neutron and neutrino experiments provide a rich arena to probe BSM physics \cite{FSNNwhitepaper2023}. 
Many of these experiments such as the Nab, pNAB, and Katrin experiments use silicon-based detector technologies due to their high energy resolution, fast timing, insensitivity to magnetic fields, and their position sensitivity \cite{Fry_2019,Baessler_2024,PSTP2024PNab,mertensNovelDetectorSystem2019,urbanCharacterizationMeasurementsTRISTAN2022}.
The University of Manitoba's high-precision mass spectrometer (``Manitoba II'') was adapted to provide a unique test station for effectively any detector-under-test (DUT).
Specifically, it was commissioned as a proton source to fill the testing requirements for the Neutron-ab (Nab) experiment, where $a$, and $b$ are the parameters being measured.

The Nab detectors have a \SI{117}{\milli\meter} diameter active area divided into 127 hexagonal pixels (each with $\sim$\SI{10}{m\meter} height). The proton beam was required to traverse the entire face of the detector, while maintaining a small ``spot size'', less than \SI{6.5}{\milli\meter\squared} area to remain within the envelope of the pixelation, enabling single pixel study without significant charge sharing between the pixels. The performance and analysis of the Nab silicon detectors using the facility is reported here \cite{nabsilicon_uofm}, while this work focuses on the development and performance of the proton source used for these studies. 
In what follows, we break down the stages within the proton source (Section \ref{sec:proton_source_components}), going through each segment in turn, and then  discuss the studies performed to understand the beam parameters under typical operation (see Section \ref{sec:proton_beam_studies}).

\begin{figure*}[ht!]
\centering
 \includegraphics[width=0.8\textwidth]{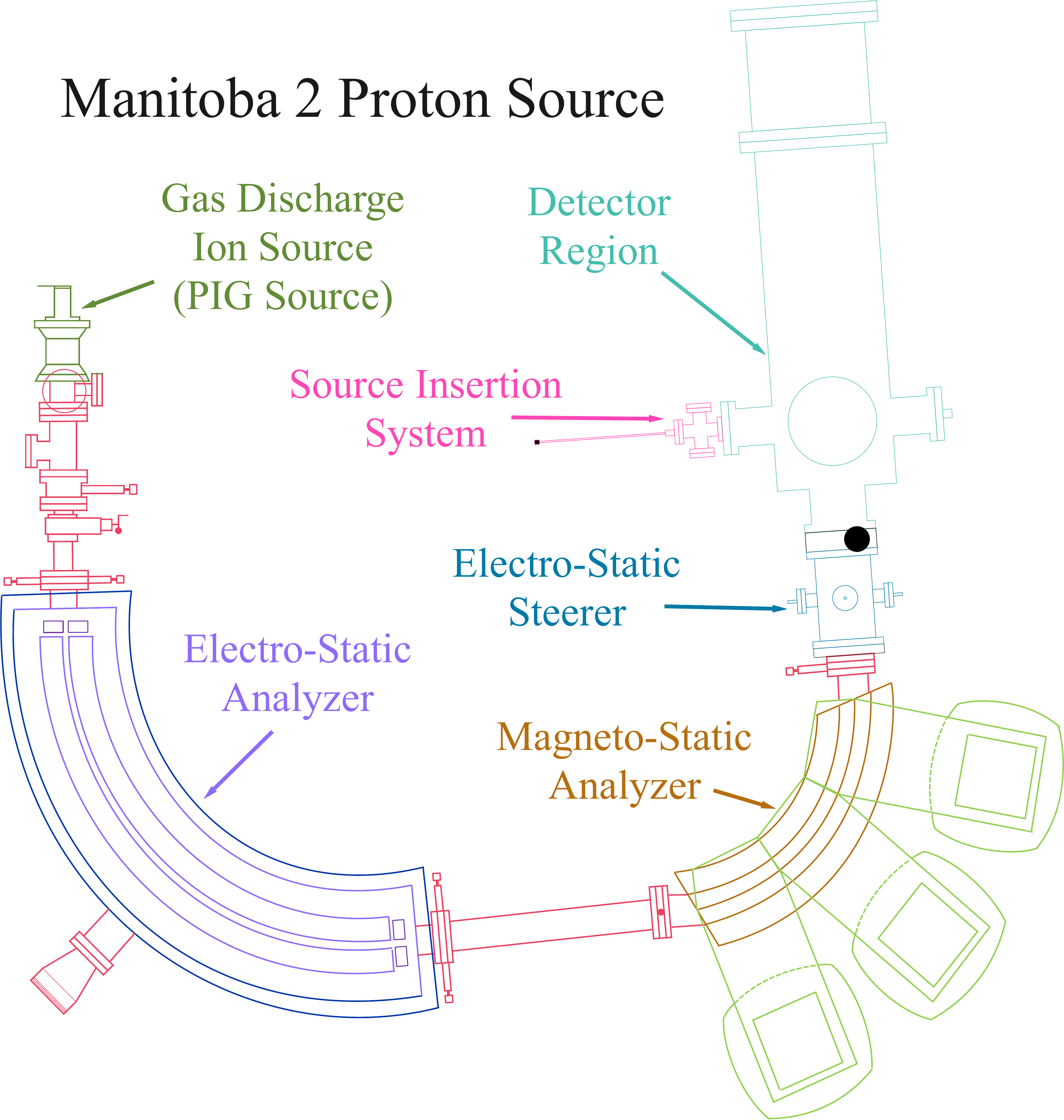}
    \caption{Top-down schematic of the Manitoba II proton source. Starting counter-clockwise from top-left, ions are born in the the gas discharge ion source (or penning ion generator (PIG), see Section \ref{sec:penning_ion_generator}), are selected based on energy in the electro static analyzer  (ESA, see Section \ref{sec:ESA}), selected based on momentum in the magneto-static analyzer (MSA see  Section \ref{sec:MSA}), pass through the steerer (see Section. \ref{sec:Steerer}), before being deposited into the detector-under-test (DUT) (see Section \ref{fig:detection_region}).}
    \label{fig:proton_source}
\end{figure*}


\section{Components of the Proton Source}
\label{sec:proton_source_components}
The Manitoba II mass spectrometer was a double-focusing (velocity and angle) high precision instrument built in 1967 for precise mass measurements of positively charged ions \cite{Barber1971}. 
After serving the atomic physics community for over forty years, it gained a second lease on life through the adaptation to a steerable low-energy ``pencil beam'' proton source for detector characterization \cite{Harrison2013}. 
Figure \ref{fig:proton_source} shows a schematic of the proton source facility after modifications. 
A gas discharge source is used to produce an ion beam, replacing the old tantalum oven-ized source (see \ref{sec:penning_ion_generator}).  
The diffusion pumps were removed from the vacuum system to reduce the risk of backward contamination to sensitive detector instrumentation in favor of more conventional turbo-molecular pumps, backed by diaphragm roughing pumps. 
The ion beam then passes through the unmodified Electro-static Analyzer (ESA) (see \ref{sec:ESA}) and Magneto-static Analyzer (MSA) (see \ref{sec:MSA}). 
An electrostatic steerer was installed after the MSA to deflect the proton beam over the face of a detector in the Detector Region. (see \ref{sec:Steerer}). 
The Detector Region includes a vacuum cross with a long cylindrical vacuum vessel downstream which accepts the \SI{1}{\meter} long Nab silicon detector, amplifier, and readout assembly \cite{Broussard2017,Fry_2019}.


\subsection{Penning Ion Generator}
\label{sec:penning_ion_generator}

\begin{figure}[ht!]
\centering
 \includegraphics[width=0.70\textwidth]{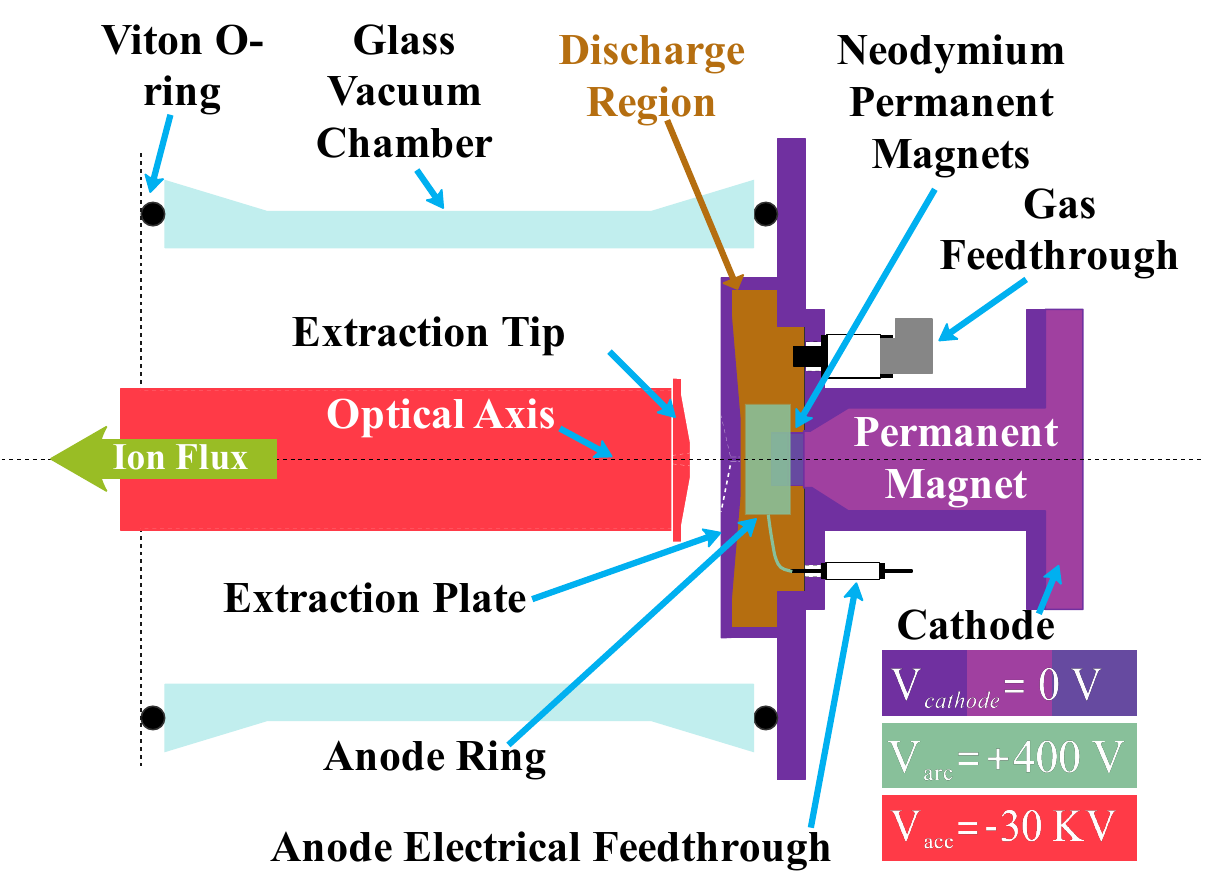}
    \caption{Cross-section of the penning ion generator used to create positive ions, including protons. A hydrogen-argon gas mixture is fed through the gas feed-through into the discharge region. A plasma is created between the extraction plate and the anode ring, ``cracking'' the hydrogen in to positive ions. The ions are then accelerated through a potential of \SI{30}{\kilo\volt} and electro-static forming plates (not shown) focus the ions into a beam.}
    \label{fig:proton_source_diagram2}
\end{figure}

A Penning ion generator (PIG) source was installed to produce protons for use in detector characterization experiments (see Figure \ref{fig:proton_source_diagram2}) \cite{Harrison2013,NickThesis}.  
A hydrogen-argon gas mixture is introduced into the high vacuum discharge region where the rate is controlled by a manual needle valve. 
The argon gas in the mixture functions as a buffer gas, limiting the mobility of free charge in the discharge vacuum region. 
This has the effect of regulating the source discharge current and ultimately the stability of the discharge, resulting in a continuous discharge.
The relative concentrations of hydrogen and argon gases in the gas mixture were empirically determined by introducing increasing quantities of hydrogen gas until the source arc current was stable at $\sim$ 1-\SI{3}{\milli\ampere}. 

In the discharge region, once an ion reaches the anode/cathode it no longer participates in the ionization process. 
A high critical-temperature permanent magnet (42\% Neodymium) is used to increase the time spent by ions ionizing other particles in the discharge region. 
Charged particles follow helical trajectories in the electromagnetic field and may even be circulated across the discharge gap multiple times before extinction depending on the ion's initial conditions and the space-charge effects of the plasma. 
Trajectories with large path lengths relative to their extinction time will have more time to ionize particles thereby increasing ion production in the discharge region. 
Ions near the extraction plate are accelerated across the \SI{30}{\kilo\volt} acceleration gap toward the extraction tip before moving to the ESA for energy analysis. 
The amount of energy imparted upon the accelerated ions depends on the type of ion.
Since protons are singly charged ions they gain \SI{30}{\kilo e\volt} of energy when accelerated through the gap, although we note that ionic $H^+_2$ and and $H_3^+$ ions are also created (see Figure \ref{fig:protonspecies}). 

\subsection{Electro-static Analyzer}
\label{sec:ESA}


The electro-static analyzer (ESA) consists of two concentric, arced plates to analyze ions according to their kinetic energy (see Figure \ref{fig:proton_source}).
The ions travel in an arced path of radius \SI{1}{\meter} due to the radial electric field produced between the plates.
This field is generated by applying a constant deflection potential $V_{d}\equiv V_{b}-V_{a}$ across the \SI{2}{c\meter} electrode gap, where the outer electrode is held at $V_{b}$ and the inner electrode is held at $V_{a}$, while requiring that $V_{b}=-V_{a}$ \cite{Liebl2008}.
Ions that possess the selected energy travel along the central circular trajectory of the ESA because the ion experiences a centripetal force $\vec{F}_{C}$ that is equal in magnitude and opposite in direction to the force of the radial electric field $\vec{F}_{E}$; thus there is no net force on ions traveling the central trajectory. 
Ions possessing energies other than the selected energy will experience a radial force which causes a them to travel a trajectory deviating from that of the central trajectory.
For \SI{30}{\kilo e\volt} ions, $V_{d}=\SI{-1200}{\volt}$.
A collimating slit at the exit of the ESA ensures only ions traveling the central circular path leave the ESA which effectively selects ions by their initial kinetic energy. 


\subsection{Magneto-static Analyzer}
\label{sec:MSA}
The magneto-static analyzer (MSA) consists of a constant current driven electromagnet with three pole pieces spanning the trajectories of the ions such that the homogeneous magnetic field is oriented normal to the incoming ion momentum.
In Figure \ref{fig:proton_source}, the magnetic field is oriented into the page.
The Lorentz force, $\vec{F}_{L}$, gives rise to an effective centripetal force, $\vec{F}_{C}$, which balances the radial component of the Lorentz force, resulting in trajectories of constant curvature (uniform circular motion):
\begin{equation}\label{MSA_eq1}
F_{C}\equiv \frac{mv_{0}^2}{r_{m}}=evB\equiv F_{L}.
\end{equation}
Where $m$ is the mass of the ion, $v$ is the ion velocity, $r_{m}$ is the radius of curvature of the circular trajectory, $e$ is the elementary charge and $B$ is the applied magnetic field. 
By substituting $p=mv$ into Equation \ref{MSA_eq1}, we find an expression that relates the ion momentum to the radius of its trajectory and the applied magnetic field given as
\begin{equation}\label{MSA_eq2}
\frac{p}{r_{m}}=eB.
\end{equation}
Since the mass of the proton is known and its energy is known from the energy analysis provided by the ESA, Equation \ref{MSA_eq2} can be used to calculate the magnetic field strength required to produce circular trajectories corresponding to the cyclotron radius $r_{m}$ for a fixed energy which coincide with the radius of curvature of the MSA (ie. $r_{m}=\SI{0.63}{\meter}$ ). 
In the special case where the ions are mono-energetic this momentum analysis is equivalent to analyzing by ion mass. 
This case is exemplified in the Manitoba II proton source since the ESA located before the MSA ensures a mono-energetic beam of ions entering the MSA. A $B= \SI{3.989e-2}{\tesla}$ is required to pass \SI{30}{\kilo e\volt} protons through the MSA.

\subsection{Proton Steerer}
\label{sec:Steerer}
\begin{figure}[ht!]
\centering
 \includegraphics[width=0.70\textwidth]{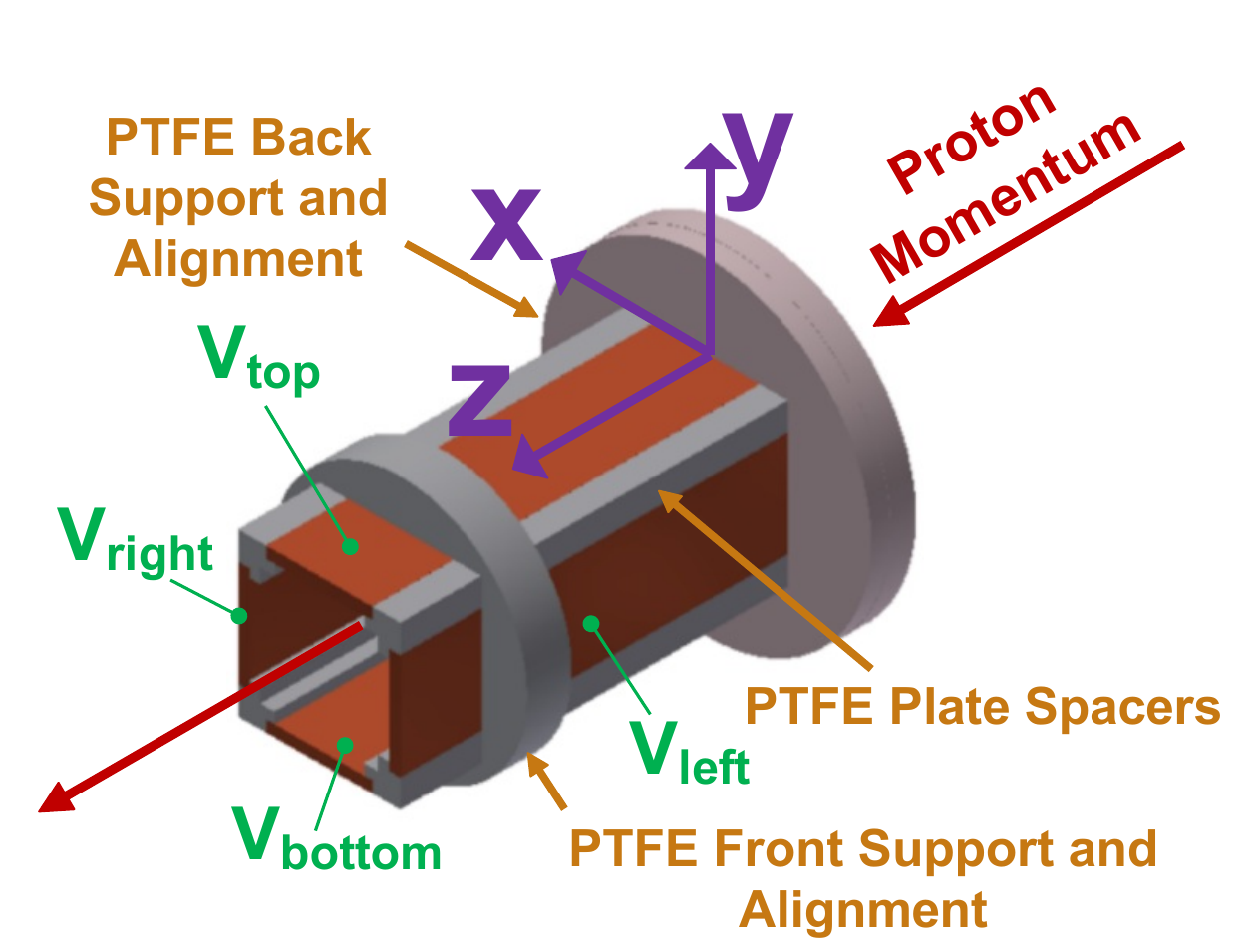}
    \caption {Model of the proton steerer assembly, showing the four plate configuration. The proton beam trajectory can be seen in red. The copper electrodes are 47.63 mm wide x 193.68 mm long. These are encased in a PTFE housing for electrical isolation. Voltages of up to \SI{\pm2}{\kilo\volt} may be individually applied to the plates to steer the beam in the desired direction.}
    \label{fig:steerer}
\end{figure}

The electro-static steerer is located in-between the MSA and the detector region. 
As shown in Figure \ref{fig:steerer}, the steerer consists of a set of 4 independent rectangular copper electrodes and is placed in a 10'' long 8'' Conflat nipple. 
Application of static potentials of up to \SI{\pm2}{\kilo\volt} can be individually applied to the parallel plates of the steerer. 
The dimensions and voltages of the steerer were determined by simulating proton trajectories through different electrode geometries using the \texttt{SIMION} ion optics simulation software such that protons would be deflected up to \SI{75}{\milli\meter} radially at the Nab silicon detector location \SI{0.48}{\meter} from exit of the electron-static steerer \cite{simion}. 

\subsection{Beam Characteristics}
\label{sec:beam_characteristics}

The proton beam can be tuned to the desired energy by adjusting the acceleration potential in the PIG.
For studies of the Nab detectors, \SI{25}{\kilo e\volt}, \SI{30}{\kilo e\volt}, and \SI{35}{\kilo e\volt} were selected.
The beam has no time structure, and instead is a constant stream of protons, with a beam current on the order of \SI{1e-18}{\ampere}.
As will be discussed below, the physical extent of the beam is quite small, less than \SI{5}{\milli\meter} in diameter, meaning that beam profile was assumed to be Gaussian.

\subsection{Detector Region}
\label{subsubsec:test_setup}
The detector region can be either fitted with a vacuum vessel to accommodate the Nab silicon detector, which is a Conflat vacuum cross featuring a 16.5'' Conflat nipple on the downstream side, or a 10'' Conflat nipple where a \SI{190}{m\meter} diameter phosphor screen or micro-channel plate (MCP) detector is placed.    
\begin{figure}[ht!]
\centering
\includegraphics[width=0.9\textwidth]{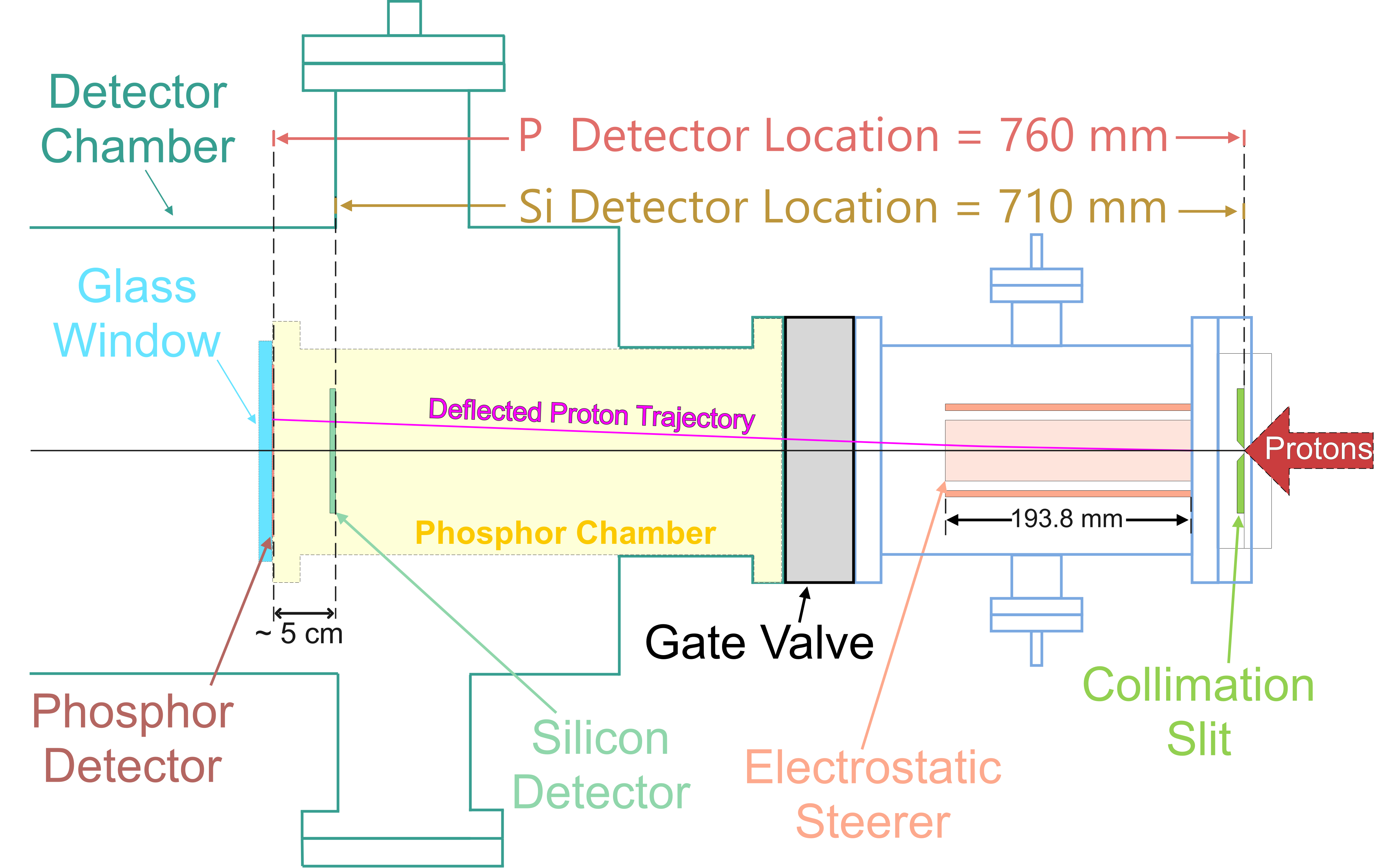}
    \caption{Top-down view of the detection region in the Manitoba II proton source. From left to right, one can see the position of the phosphor screen and supporting vacuum extension (cyan and yellow, respectively) - this configuration was only installed for a short time to study the beam movement (see Sec. \ref{subsubsec:phosphor_screen}). After this was removed, the detection chamber was installed. This included the silicon detector, and calibration sources. During this campaign, a pencil beam of protons (emerging from the collimation slit) could be used to probe individual segments of the silicon detector, via the electro-static steerer (Sec. \ref{sec:Steerer}).}
    \label{fig:detection_region}
\end{figure}
Figure \ref{fig:detection_region} shows a cross-sectional view of the Detector Region along with the vacuum beam pipe containing the steerer.  
Here one can see the phosphor screen is \SI{5}{\centi\meter} further downstream of where the Nab Si detector would be located. 
The vacuum vessel used for the Nab silicon detector was fitted with a turbo-molecular pump backed by a diaphragm pump, as well as feed-through for Si detector temperature sensor readout, residual gas analysis, and a retractable rod with radioactive sources on the end. 
This vessel was maintained at pressures at or below \SI{5e-7}{Torr}. 

\section{Proton Beam Studies}
\label{sec:proton_beam_studies}
We present three major studies that define the characteristics of the proton beam in the detector region: the proton energy resolution, proton spot size at various radii as measured by the phosphor screen, and a proton spot size measurement made with the Nab silicon detector.
For the proton energy resolution and the phosphor screen study, the PIG source was optimized for highest arc current, which maximizes proton rate.
For the silicon detector spot size study, electrical noise necessitated reducing this to improve the signal-to-noise ratio.
This meant a proton rate of $\sim$\SI{10}{counts\per\second}.

\subsection{Proton Energy Resolution}
\label{sec:proton_energy_resolution}

\begin{figure}[ht!]
\centering
 \includegraphics[width=0.5\textwidth]{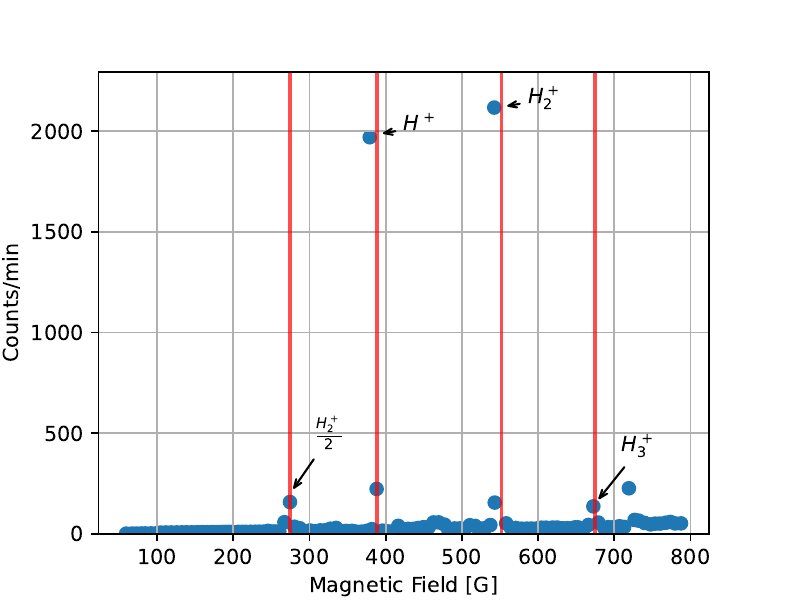}
    \caption{A Micro-channel plate (MCP) detector measures counts per minute with pencil beams of various hydrogen species passing through both analysis sectors while varying the magnetic field in the MSA. The solid vertical lines are calculated using a \SI{27.245}{\kilo\volt} acceleration voltage and Equation \ref{MSA_eq1}. This acceleration voltage can be varied within a range of stability --- for this test \SI{27.245}{\kilo\volt} was the highest stable voltage, however improvements were made to achieve up to \SI{35}{\kilo\volt}. Also note that the count rate is highly dependent on the PIG configuration, and the noise in the detector used \cite{Harrison2013}.}
    \label{fig:protonspecies}
\end{figure}

These studies show that the we are indeed able to select protons and determine the energy resolution of those protons after they exit the MSA.  
Here a \SI{30}{k\volt} accelerating potential is used with an off-the-shelf micro-channel plate detector (Photonis) placed at the center of the beam pipe after the MSA. 
Figure \ref{fig:protonspecies} shows the counts vs magnetic field of the MSA with several mass peaks identified, where separation distance in magnetic field units identifies protons, $H_2$, and $H_3$  ions.
Solving for $B$ in Equation \ref{MSA_eq1}, and substituting the kinetic energy, $T = mv^2/2$, gives
\begin{equation}\label{eq:magField}
    B = \frac{\sqrt{2Tm}}{er_m}.
\end{equation}
For the various ions, the magnetic field strengths can then be found.
These are represented by the red vertical lines in Figure \ref{fig:protonspecies}.

\begin{figure}[ht!]
\centering
 \includegraphics[width=0.5\textwidth]{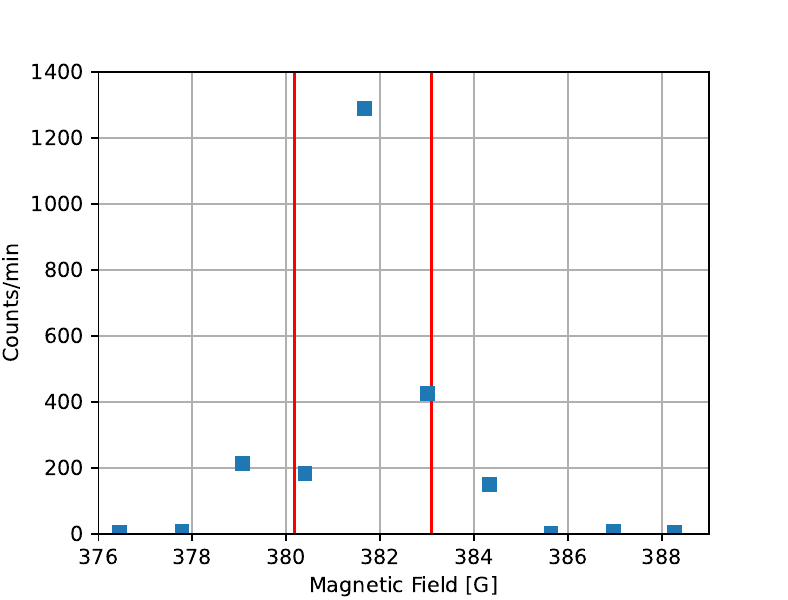}
    \caption{Proton counts as a function of magnetic field strength in the MSA to resolve the $H^+$ peak. The red vertical lines represent \SI{2}{G} FWHM. This data translates to a proton energy resolution of \SI{300}{e\volt} via Equation \ref{scur}. This is much lower than the energy resolution of the Nab silicon detectors, allowing a precise confirmation of the Nab detector performance \cite{Broussard2017}.}
    \label{fig:proton_magscan}
\end{figure}

Figure \ref{fig:proton_magscan} shows a finer scan in magnetic field across the $H^+$ peak, where the full-width at half-maximum (FWHM) of the peak is $\sim$ \SI{2}{Gauss} (represented by the red vertical lines).
Converting equation \ref{MSA_eq2} into kinetic energy $T$, we find that
\begin{equation}\label{scur}
T=\frac{e^2B^2r_m^2}{2m} \Rightarrow \frac{\delta T}{T} = \frac{\delta B^2}{B}.
\end{equation}
Thus the error in our proton energy is $\sim$\SI{300}{e\volt} FWHM, which is generally much lower than the energy resolution of the Nab silicon detectors \cite{Broussard2017}.

\subsection{Phosphor Screen Study}
\label{subsubsec:phosphor_screen}
To ensure that the beam spot size in the detector plane is smaller than a single Nab detector pixel even after deflection, an in-situ measurement of the beam spot shape was conducted using photography of a phosphorescent detector near the proposed location of the Nab detector on the proton source beamline. 
A phosphor detector was purchased from ProxiVision GmbH\textsuperscript{\texttt{TM}} and consists of a P43 ($Gd_2O_2S$:$Tb$) phosphor coating on a \SI{200}{\milli\meter} diameter borosilicate glass substrate which emits light in the visible green spectrum at $\sim \SI{545}{\nano\meter}$ making it photographically compatible with a typical digital single lens reflex (SLR) camera. 
The utility of a phosphor screen detector for beam spot measurements stems from its high positional resolution and minimal electronics readout. 
In our implementation of the phosphor detector, the positional resolution was limited by the camera that photographs the screen.

\begin{figure}[ht!]
\centering
\includegraphics[width=0.5\textwidth]{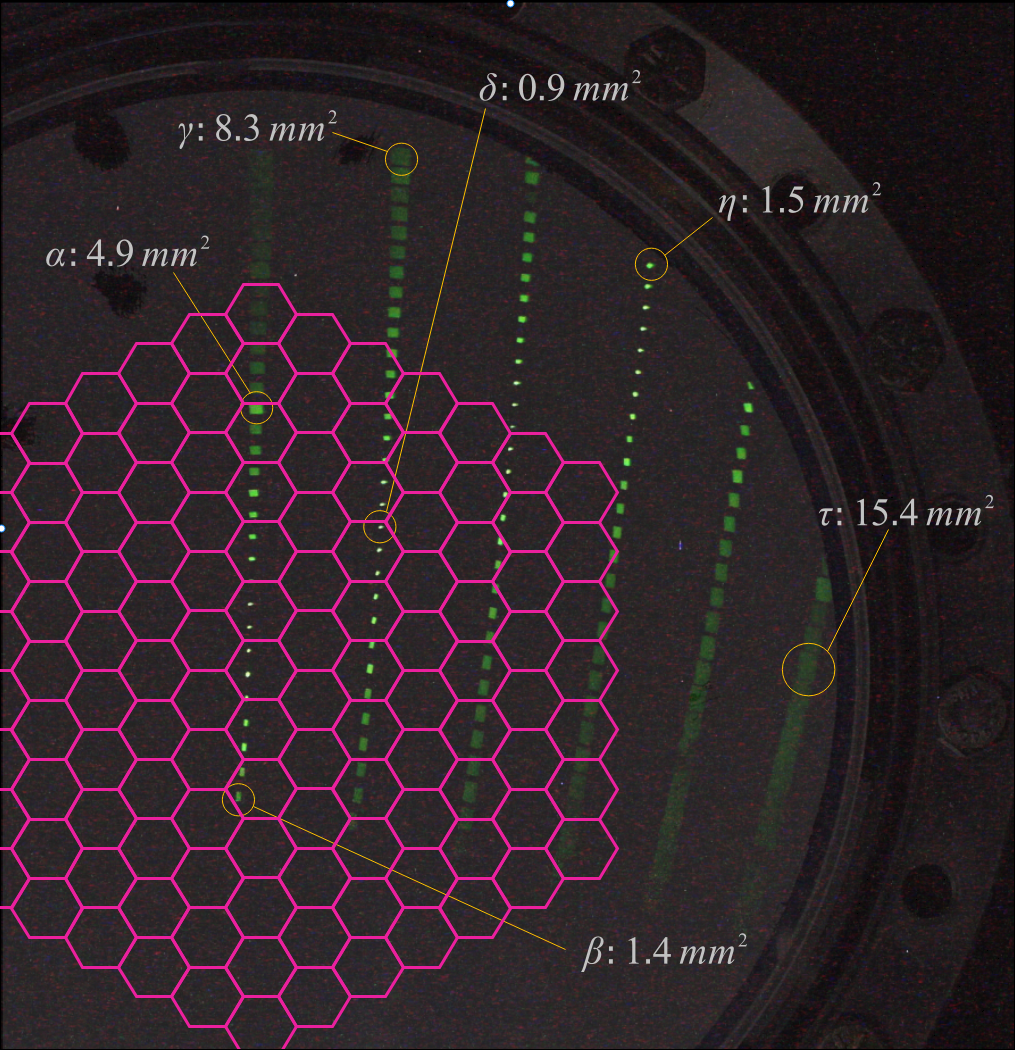}
    \caption{Composite photograph of proton beam spot at an array of deflection voltages. The pixilated Nab Si detector is also overlayed to show that the proton spots sizes are well within a pixel.}
    \label{phosphor_photo2}
\end{figure}

Beam spot sizes on the order of \SI{1}{\milli\meter\squared} were easily resolvable in low light conditions. 
An example of a set of beam spot measurements is shown in a composite photograph in Figure \ref{phosphor_photo2}. 
In this measurement, the collimation slit is made as small as low light photography would permit the extraction of beam spot images near the circumference of the active area of the phosphor screen. 
For this, the collimation slit separation of $\sim \SI{2}{\milli\meter}$, the beam spot was photographed at various deflection voltages, one spot at a time. 
All beam spot photographs were then combined into a single image to produce a beam spot array.
This array which was then superimposed over an image of the detector in full light conditions to show the scale of the beam spot size with respect to the sizes of the phosphor detector and vacuum flange holding it. 
Using these known values an conservative estimate of the width and height of each beam spot was determined. 
The beam spot size is shown to vary with respect to deflection voltage and beam spot position as expected due to the finite beam interacting with non-uniform electric fields in the steerer. 
The largest beam spots are observed at extreme positions along the cardinal directions. 
The smallest beam spots are observed at small deflections and/or at positions along the ordinal directions. 
Examples of extreme spot sizes are shown in Figure \ref{phosphor_photo2} by $\delta$-spot with an area of \SI{0.9}{\milli\meter\squared} and $\tau$-spot with an area of \SI{15.36}{\milli\meter\squared} both of which are smaller than the Nab detector pixel area of \SI{70}{\milli\meter\squared}.
 
\subsection{Proton Beam Spot Size Determination with a Segmented Silicon Diode Detector}
\label{subsec:Proton_spot_size_study}

Motivated as a complement to the phosphor screen study mentioned above, further confirmation of the beam size was undertaken using one of the Nab segmented silicon diode detectors. 

\begin{figure}[ht!]
    \centering
    \includegraphics[width=0.5\textwidth]{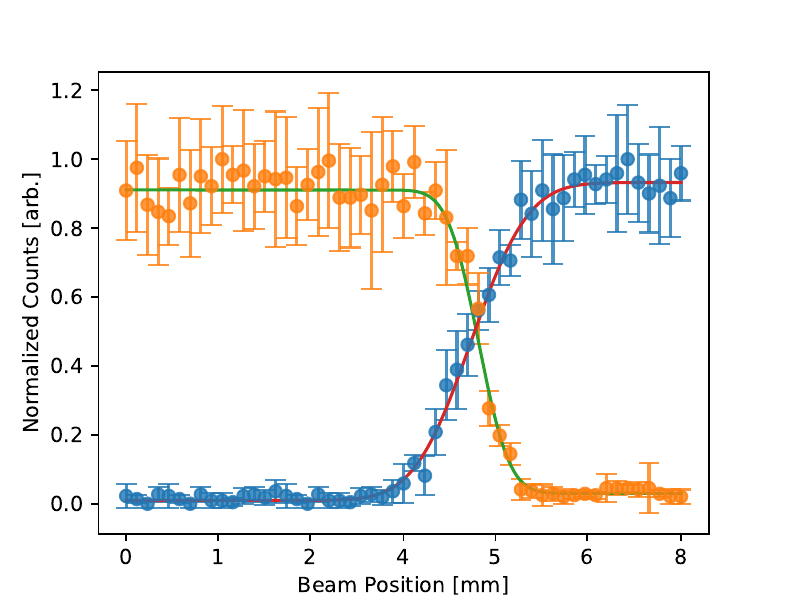}
    \caption{Normalized counts from a pencil beam of \SI{30}{\kilo e\volt} protons moving across the boundary between two silicon detector elements.} 
    \label{fig:beam_diameter}
\end{figure}

The process employed the boundary between two detector segments.
In Figure \ref{phosphor_photo2}, the central pixel is labeled as ``zeroth ring'', whereas the pixels uesd in this study were located in the third ring.
By using a single-channel analyzer to capture the number of events in a gated time window from the segment of interest, count rate as a function of beam-position was recorded as the beam was moved off of one segment and on to the other.  
The count rate rolled off with the approximate form of a combination of error functions, given by 
\begin{equation}\label{eqn:errFunc}
     f(x) = \frac{a}{2}\left[b\pm\text{erf}\left(\frac{x-\mu}{\sqrt{2}\sigma}\right)\right].
\end{equation}

Because the beam is steered with a capacitive steerer (Section \ref{sec:Steerer}), the beam deflection voltage is the input parameter. The beam was driven across the entire detector segment of known dimension in order to convert deflection voltage in to spatial units. 
This can be seen in Figure \ref{fig:beam_diameter}.
Fitting the count rate as a function of beam position and comparing to the physical dimensions of the detector segment gives the conversion factor.

By extracting the half-width at half maximum (HWHM), the diameter was taken to be 
\begin{equation}\label{eq:proton_beam_diameter}
    D_{beam} = 2\times \text{HWHM} =  2\sigma\sqrt{2\ln(2)}.
\end{equation}
The largest contributions to the uncertainty in this value come from performing this study off of the optical axis, and due to the geometry of the steerer, this necessitated a correction to the horizontal beam deflection while driving the beam vertically (see Sec. \ref{sec:Steerer}). 
Further, the power supplies had relatively coarse (\SI{1}{\volt}) step sizes.

The average values obtained were \SI{3.1 \pm 0.2}{\milli\meter\squared}, which agrees with the phosphor screen study.
In light of this, we meet the requirement of constraining the beam spot size to within the footprint of a single pixel on the Nab detectors.
This was confirmed by steering the beam across one of the pixels in two directions, and empirically determining a steering voltage based on the count rate in that pixel as well as its six neighbours.

\section{Conclusion}
\label{sec:conclusion}

The Manitoba II spectrometer was modified with a \SI{30}{\kilo e\volt} accelerating potential penning trap ion source and a electro static steerer with the goal of characterizing the 117 mm diameter, 127 pixel Nab silicon detectors. 
The spectrometer easily isolates the \SI{30}{ke\volt} protons generated from the penning ion generator with an energy resolution of \SI{300}{e\volt} FHWM. 
During the testing of the Nab detectors, we ran at \SI{10}{protons\per\second}, to reduce electrical noise in the data acquisition system. 
To evaluate the proton spot size, a large diameter phosphor screen was employed where most spot sizes in the area of interest were \SI{1}{m\meter^2} area with a maximum area of \SI{5}{\milli\meter\squared}. 
In addition the spot size was evaluated by scanning over a pixel on the Nab silicon detector which determined a spot size diameter of \SI{3.1}{\milli\meter\squared}.
This facility played a major role in the characterization of one of the large-area silicon detectors used in the Nab experiment, meeting the requirements of segment-by-segment calibration and study.

\section{Acknowledgments}
We acknowledge the support of the Natural Sciences and Engineering Research Council of Canada (NSERC), [contracts SAPPJ-2019-00043, SAPPJ-2022-00024, SAPPJ-2025-00045]. This research was sponsored in part by the U.S. Department of Energy (DOE), Office of Science, Office of Nuclear Physics [contract DE-AC05-00OR22725 and 9233218CNA000001 under proposal LANLEEDM] and in part by the U.S. DOE, Office of Science, Office of Workforce Development for Teachers and Scientists, Office of Science Graduate Student Research program.












\bibliographystyle{ieeetr}
\bibliography{refs.bib}
\end{document}